\documentclass[aps,prl,reprint,showkeys,noshowpacs,hyphens,superscriptaddress,floatfix, nofootinbib]{revtex4-1}
\usepackage{amsmath, amsfonts, amssymb, bm,gensymb,harpoon,textcomp,ulem,comment,graphicx,dcolumn,blindtext}
\DeclareRobustCommand{\orderof}{\ensuremath{\mathcal{O}}}
\usepackage{hyperref}
\hypersetup{colorlinks=true,urlcolor=blue,citecolor=blue,linkcolor=blue} 
\usepackage{geometry}		
\usepackage{xcolor}
\definecolor{commentblue}{rgb}{0,0,1}

\begin{document}

\preprint{APS/123-QED}

\title{Probing hydrodynamic fluctuation-induced forces with an oscillating robot}

\author{Steven W. Tarr}
\affiliation{School of Physics, Georgia Institute of Technology, 837 State Street, Atlanta, Georgia 30332, USA}
\author{Joseph S. Brunner}
\affiliation{School of Physics, Georgia Institute of Technology, 837 State Street, Atlanta, Georgia 30332, USA}
\affiliation{Department of Radiation Medicine, University of Kentucky, 800 Rose Street, Lexington, Kentucky 40536, USA}
\author{Daniel Soto}
\affiliation{School of Physics, Georgia Institute of Technology, 837 State Street, Atlanta, Georgia 30332, USA}
\author{Daniel I. Goldman}
\affiliation{School of Physics, Georgia Institute of Technology, 837 State Street, Atlanta, Georgia 30332, USA}
\email{daniel.goldman@physics.gatech.edu}

\date{\today}

\begin{abstract}
We study the dynamics of an oscillating, free-floating robot that generates radially expanding gravity capillary waves at a fluid surface. In open water, the device does not self-propel; near a rigid boundary, it can be attracted or repelled. Visualization of the wave field dynamics reveals that when near a boundary, a complex interference of generated and reflected waves induces a wave amplitude fluctuation asymmetry. Attraction increases as wave frequency increases or robot-boundary separation decreases. Theory on confined gravity-capillary wave radiation dynamics developed by Hocking in the 1980s captures the observed parameter dependence due to these “Hocking fields.” The flexibility of the robophysical system allows detailed characterization and analysis of locally generated nonequilibrium fluctuation-induced forces~\cite{kardar1999}.

\end{abstract}

\maketitle

Forces mediated by steady-state fluctuations in fields are well studied in diverse systems~\cite{casimir1948influence,casimir1948attraction,li1991,kardar1999,lee2017fluctuation}. In confinement, emergent wave field asymmetries can produce nonzero, net fluctuation-induced forces on boundaries; such forces are observed across scales, from the quantum mechanical vacuum ~\cite{casimir1948influence,casimir1948attraction,kardar1999,klimchitskaya2009casimir,woods2016materials} to fluids~\cite{li1991,boersma1996maritime,denardo2009water,harris2013wavelike,pucci2016nonspecular,lee2017fluctuation,saenz2021emergent}.  In quantum mechanics, the Casimir effect demonstrates that nearby neutral plates confine and modify zero-point-energy wave fields, often yielding attraction~\cite{casimir1948influence,casimir1948attraction,kardar1999,klimchitskaya2009casimir,woods2016materials}.  In driven fluid systems, boundaries generate an analogous downsampling of surface wave modes called the ``maritime Casimir effect''~\cite{boersma1996maritime,lee2017fluctuation}. The downsampled modes reduce the radiation pressure between objects at the fluid surface and can be observed as reduced amplitude waves~\cite{boersma1996maritime,denardo2009water,parra2014casimir,lee2017fluctuation}.

More recently, researchers studying nonequilibrium fluctuation-induced forces have uncovered a variety of Casimir-like phenomena that demonstrate long-range attraction and repulsion in diverse systems including complex fluids, fluid membranes, and vibrofluidized granular media~\cite{najafi2004,buenzli2008,hanke2013,parra2014casimir,ray2014casimir,ni2015tunable,aminov2015}. Such systems sustain additional effects owed to their nonequilibrium dynamics including generic power law correlations~\cite{aminov2015}, violations of Newton's Third Law~\cite{buenzli2008}, and migration toward colder regions~\cite{najafi2004}. Indeed, the past 30 years have generated expansive literature on Casimir and Casimir-like forces. However, to our knowledge, these forces have not been leveraged for self-propulsion. The capacity for locomotion stemming from symmetric momentum generation is novel and stands in contrast to typical asymmetric inertial self-propulsion (e.g., body bending~\cite{purcell1977life,maladen2009undulatory,kantsler2013ciliary,diaz2022water,chong2023self}, wave expulsion~\cite{roh2019honeybees,lee2019milli,rhee2022surferbot,benham2022gunwale}, spinning propellers~\cite{gerr1989propeller}).

\begin{figure}[t!]
    \includegraphics[width=0.5\textwidth]{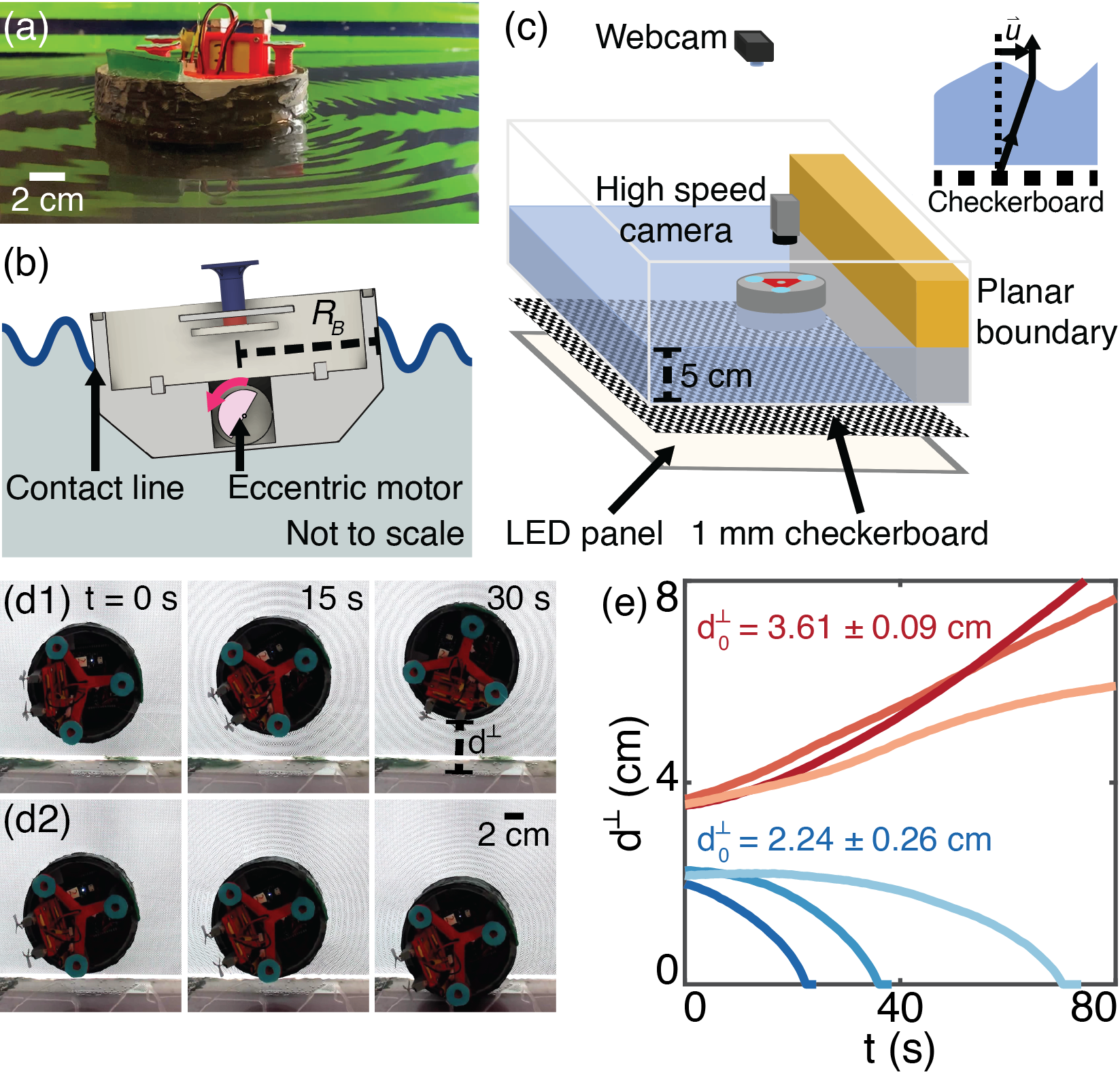}
    \centering
    \caption{\textbf{Wave-generating robot boat.} (a) Photo of boat generating 17.1 Hz waves. (b) Schematic of the eccentric motor vibrating the boat to generate waves; propellers shown in (a) are not used in this study and thus omitted in (b). (c) Diagram of the tank wherein all experiments were performed. A backlit checkerboard enables Fast Checkerboard Demodulation for spatiotemporal surface reconstruction~\cite{wildeman2018real}. (\textit{Inset}) Fast Checkerboard Demodulation determines fluid surface height using the instantaneous distortion of a checkerboard by surface perturbations. (d1-2) Time series of repulsion from (17.1 Hz) and attraction toward (33.5 Hz) wall, respectively. (e) Evolution of perpendicular hull-wall distance for repeated repulsive and attractive trials at 17.1 Hz and two different initial distances.}
    \label{fig:IntroToBoat&Apparatus}
\end{figure}

Here we introduce a system that allows not only for convenient creation and visualization of non-equilibrium fluctuation-induced forces using surface waves but also for probing a new regime where the agents subject to fluctuation-induced forces are themselves producing the requisite fluctuations. In doing so, we also discover that self-propulsion can be induced in a free-floating, oscillating robophysical system that does not directly generate asymmetric momentum transport. Symmetrically propagating waves undergo a complex interference when reflected at a boundary, breaking symmetry and generating propulsive radiation forces. We probe the dynamics with a custom-developed robot and map radiation forces as both oscillation frequency and confinement distance vary. Confinement on one side leads to a modification in wave field amplitude, and the dependence of the consequent radiation force on oscillation frequency can be quantitatively explained by theory for gravity-capillary waves developed by Hocking~\cite{hocking1987reflection,hocking1987damping,hocking1988capillary}. Further, we demonstrate the capacity for fluctuation-induced forces in systems with monochromatic fluctuations as opposed to the typical noisy spectra~\cite{casimir1948influence,casimir1948attraction,li1991,kardar1999,cattuto2006,denardo2009water,klimchitskaya2009casimir,parra2014casimir,ray2014casimir,aminov2015,lee2017fluctuation}. Given the importance of the seldom-studied generation and reflection properties of gravity-capillary waves to the boat’s locomotion, we refer to these confined, asymmetric wave fields as ``Hocking fields.''

\textit{Apparatus \& fundamental behaviors}~--~The robotic boat (total mass $m = 368$ g) consists of a circularly symmetric hull of radius $R_B = 6$ cm, a custom circuit board, two fan motors (uxcell Coreless Micro Motor 412), and an eccentric motor (Vybronics Inc. Cylindrical Vibration Motor VJQ24-35K270B). The boat’s hull was 3D printed in Polymaker\texttrademark~PLA and waterproofed with marine epoxy. All electronics and batteries were mounted onboard, and additional weights were added such that a free-floating boat at rest is level to within $1\degree$. We mounted the eccentric motor beneath the electronics; when enabled, the motor vibrates the boat with power-dependent frequency $\omega$ primarily along the fore-aft axis (roll) with minimal vertical motion or induced surface currents. Beyond $\omega = 20$ Hz, the vibration tends toward roll amplitude $0.15\degree \pm 0.02\degree$, pitch (left-right axis) amplitude $0.05\degree \pm 0.01\degree$, and vertical oscillation amplitude $0.09 \pm 0.02$ mm (see SI). The result is a left-right and fore-aft symmetric, radially emanated, monochromatic wave train of wavelength $\lambda(\omega)$ traveling along the fluid surface (Fig.~\ref{fig:IntroToBoat&Apparatus}(a-b), Movie S1).

Because of the symmetries of the emitted waves, a boat placed far from boundaries experiences no net radiation force $F_W$. Upon breaking symmetry by approaching a boundary, $F_W$ becomes nonzero, and the boat self-propels (Figs.~\ref{fig:IntroToBoat&Apparatus}(d1-2), Movie S2). We observe both repulsive and attractive behaviors (Fig.~\ref{fig:IntroToBoat&Apparatus}(e)), with repulsion occurring more weakly such that it is often indistinguishable from noise.

To probe these dynamics, we placed the boat near a rigid acrylic planar boundary extending from the floor above the water (61 cm long, 30 cm tall, vertical to within $1\degree$), varied both $\omega$ and initial hull-boundary distance $d^\perp_0$, and allowed the boat to move freely in response to $F_W$. Though we were unable to prescribe wave amplitude $A$ independently from $\omega$, we expect it to affect $F_W$ in accord with established theory on the energy of surface waves~\cite{lamb1924hydrodynamics,longuet1964radiation}. We chose a wall with length $\ell \gg R_B,\lambda$ such that we may treat our system as quasi-1D and study the boat's motion along the axis normal to the wall. Any observed parallel motion had no clear bias. For all experiments, we programmed a motor controller to ramp the eccentric motor up linearly to the target frequency over $10$ s to minimize transients.

\begin{figure}[t!]
     \includegraphics[width=0.5\textwidth]{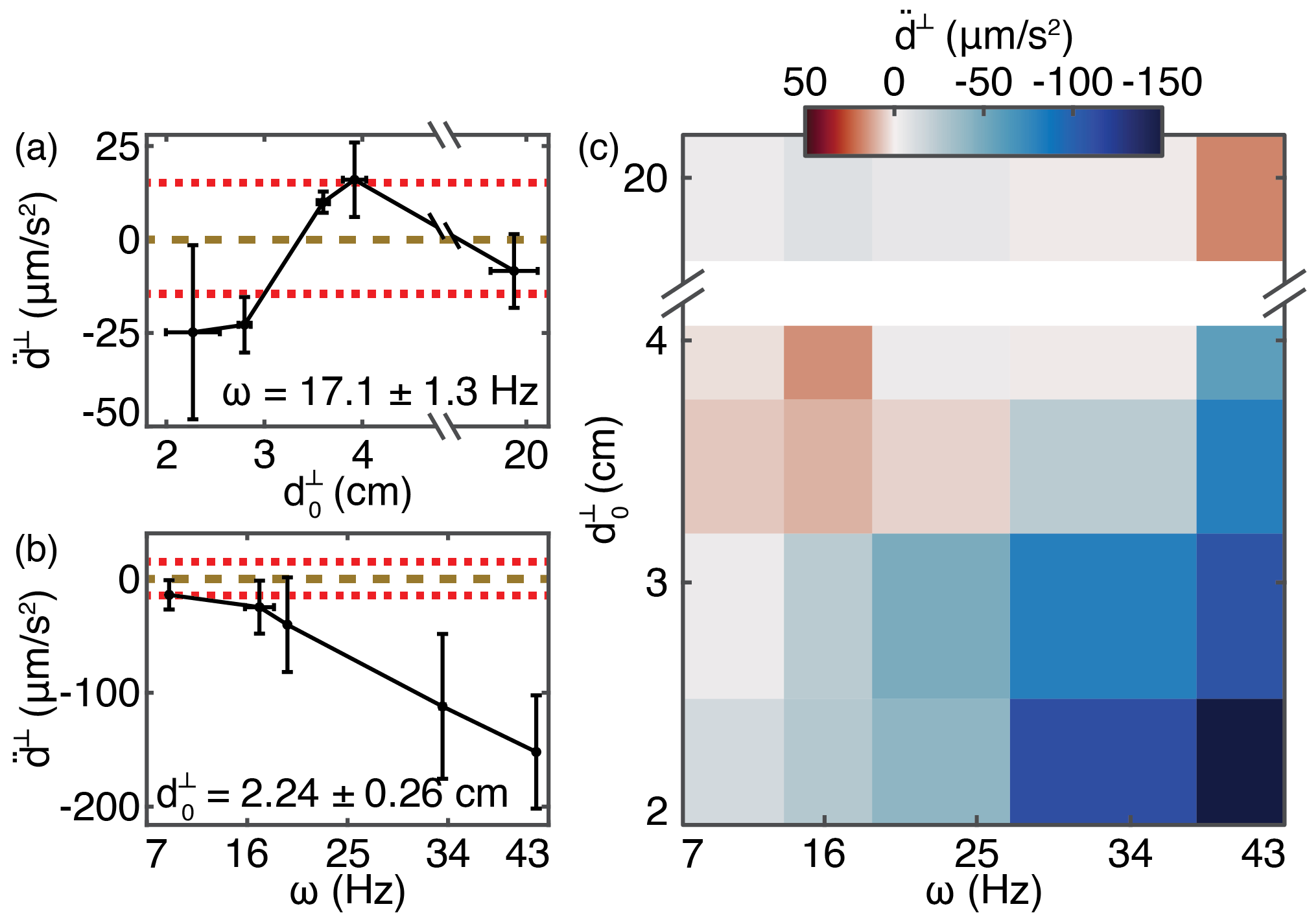}
   
    \centering
    \caption{\textbf{Wave-generating boat experiences attraction and repulsion near boundaries.} (a-b) $\ddot{d}^\perp$ versus $d^\perp_{0}$ and $\omega$. Red dotted lines denote the system's noise interval determined by behavior far from boundaries. Simultaneous dependence on both parameters is shown in (c), where each box corresponds to the average of 5 trials.}
    \label{fig:Attraction&Repulsion}
\end{figure}

We recorded images of trials with a Logitech C920 webcam at 30 FPS and tracked the boat's lateral motion with color-thresholding code in MATLAB. We extracted the boat's perpendicular acceleration $\ddot{d}^\perp$ by fitting a quadratic equation to the position-time data prior to any drag-induced inflection point. We observed an increasingly attractive force with decreasing $d^\perp_0$ and increasing $\omega$ (Figs.~\ref{fig:Attraction&Repulsion}(a-b)). During some trials with high $d^\perp_0$ and low $\omega$, a lightly repulsive $F_W$ emerged. For particularly high $d^\perp_0$, the boat was considered to be ``far from boundaries''; the wave field symmetry was restored and the boat experienced a near-zero $F_W$. We refer to the threshold distance separating the attractive and repulsive regimes as $d^\perp_T(\omega)$ (see SI).

\begin{figure}[t!]
    \includegraphics[width=0.5\textwidth]{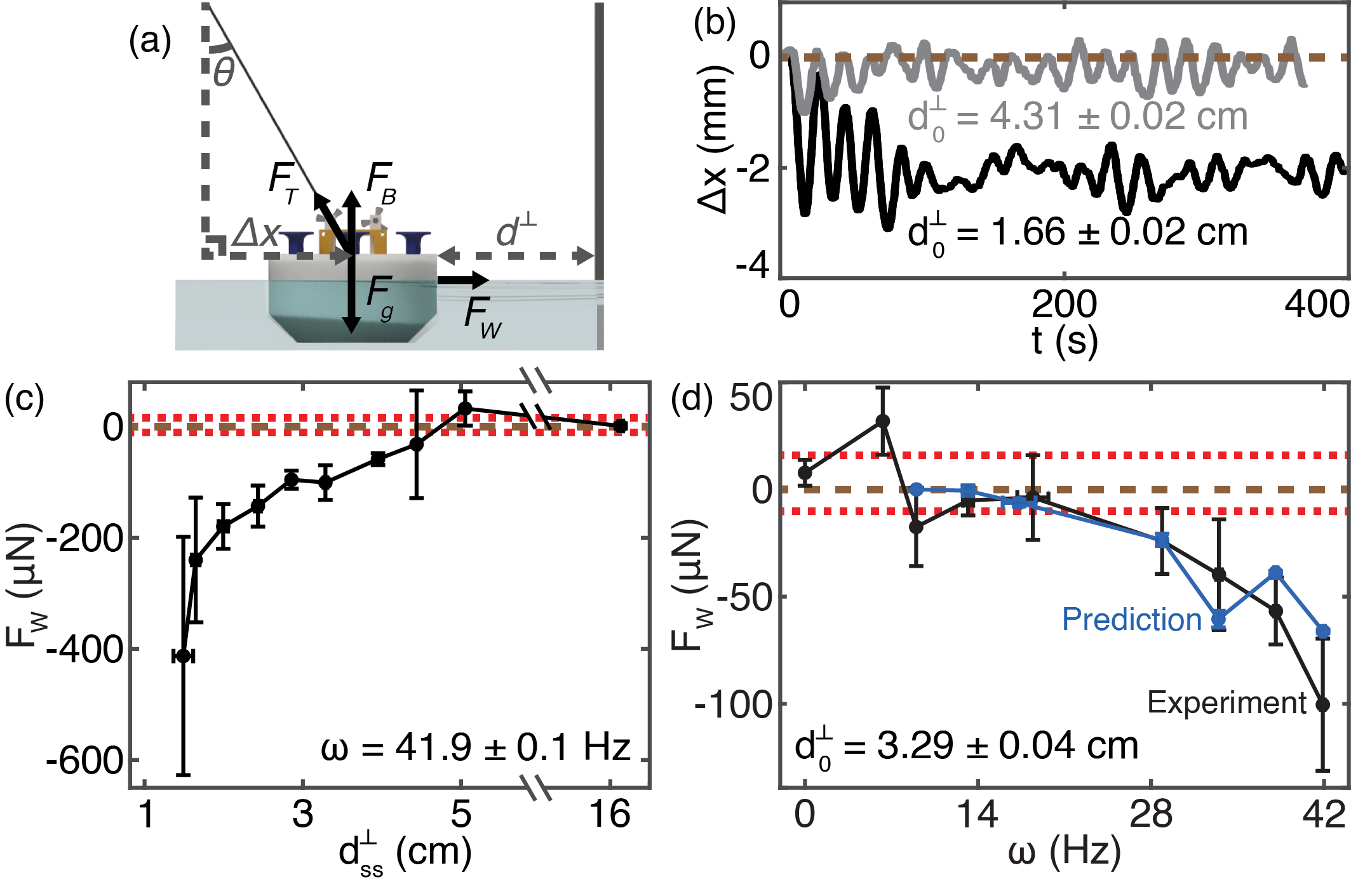}
   
    \centering
    \caption{\textbf{Measurement of near-boundary propulsive force and its parameter dependence.} (a) Diagram for pendulum experiments used to directly measure $F_W$. (b) Archetypal boat displacement plots for pendulum experiments at 33.5 Hz and two different initial distances. Oscillations are attributed to the interplay between $F_W$ and $F_T$. (c-d) $F_W$ versus $d^\perp_\text{ss}$ and $\omega$. Red dotted lines denote the system's noise interval determined by behavior far from boundaries. Blue line in (d) indicates theoretical prediction using measurements in Fig.~\ref{fig:Waves&Imaging} and Eq.~\eqref{GCRadiationForce}.}
    \label{fig:PendulumExpts}
\end{figure}

\textit{Direct measurement of wave force}~--~Having observed $\orderof(\ddot{d}^\perp) \leq 10^{2}$ $\mu$m/$s^2$ across all tested initial conditions, we sought to isolate $F_W$ from any transient effects (e.g., viscous~\cite{danov2000viscous} and wave~\cite{burghelea2001onset,diaz2022water} drag, inertia~\cite{scholz2018inertial}) that could dampen the system's evolution and result in such a minuscule acceleration. We investigated $F_W$ alone by restricting the boat's motion to that of a simple pendulum without impeding vibration (Fig.~\ref{fig:PendulumExpts}(a)), a method similarly employed to quantify water wave analog Casimir forces~\cite{denardo2009water}. The boat was affixed along its central axis $1.3$ cm above the waterline to a thin fishing line of length $L = 1.4$ m via a bowline knot. We calibrated the line such that when the pendulum angle $\theta$ was zero, the tension force $F_T$ too was zero. For nonzero $F_W$, the boat's resultant displacement $\Delta x$ caused $F_T$ to increase until reaching force balance (Fig.~\ref{fig:PendulumExpts}(b)). We measured $\Delta x$ for a variety of $\omega$ ($0$-$42$ Hz) and $d^\perp_0$ ($1.9$-$3.8$ cm) and observed typical values within $0$-$3$ mm. Since $L \gg \Delta x$, we assume the boat undergoes negligible vertical displacement~\cite{NegligibleVertDispNote}.

By measuring $\Delta x$ in steady state, we can estimate the perpendicular wave force $\overrightharp{F}_W = (m-\rho V)g\Delta\overrightharp{x}/L$, where $\rho$ is the fluid density and $V$ is the liquid volume displaced by the boat. We plotted $F_W$ as a function of the steady-state hull-wall distance $d^\perp_\text{ss}$ and $\omega$ (Figs.~\ref{fig:PendulumExpts}(c-d)). We include a heatmap of all force data in Fig.~S4. As expected, the qualitative behavior of $F_W$ closely resembles that of the acceleration, with increasing attraction below, light repulsion near, and near-zero effects above $d^\perp_T(\omega)$. Despite the removal of transient effects, attractive and repulsive forces remained small, respectively demonstrating $\orderof(F_W) \in [10^{1},10^{2}]$ and $[10^{0},10^{1}]$ $\mu$N. We note that most measurements with $d^\perp_\text{ss} < d^\perp_T$ fall outside the experimental noise interval $2.9 \pm 13.1$ $\mu$N.

\begin{figure}[t!]
    \includegraphics[width=0.5\textwidth]{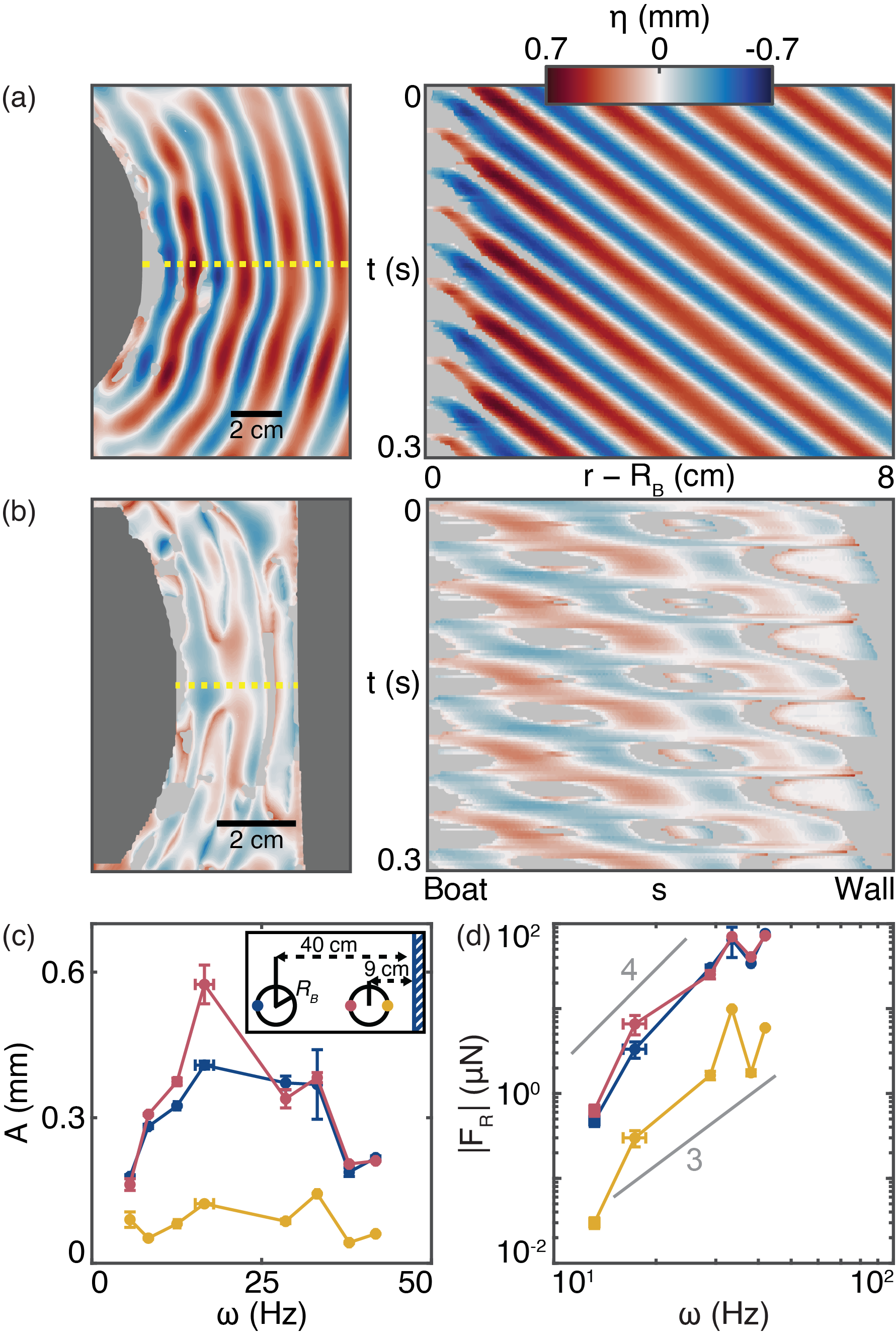}
    \centering
    \caption{\textbf{Visualization and quantification of near-robot gravity-capillary wave fields.} (a-b) Reconstructions of 17.1 Hz waves far from and near a boundary, respectively, with space-time heatmaps corresponding to dotted yellow lines. $\eta(t,\overrightharp{r}$) describes the free surface height with respect to $h_\text{rest}$. Dark gray regions were occupied by solid objects (e.g., boat, wall). Light gray regions were deemed unreconstructable (see SI). (c) Fast Checkerboard Demodulation measurements reveal the net field near the wall to have reduced $A(\omega)$. (d) Radiation forces on boat sides due to emitted gravity-capillary waves as predicted by Eq.~\eqref{GCRadiationForce} and panel (c). Solid gray lines denote scaling of $\omega^3$ and $\omega^4$.}
    \label{fig:Waves&Imaging}
\end{figure}

\textit{Surface wave measurements}~--~To better understand the role of the emanated waves in generating a locomotive force, we employed the synthetic~\cite{SyntheticNote} Schlieren visualization technique Fast Checkerboard Demodulation~\cite{wildeman2018real} (see SI) to obtain quantitative measurements of the wave field (Figs.~\ref{fig:IntroToBoat&Apparatus}(c),~\ref{fig:Waves&Imaging}(a-b), Movie S3). For optimal visualization quality, we minimized the water's depth to $h_\text{rest} = 5$ cm for all experiments. Imaging was performed with a high speed camera (AOS X-PRI) at 500 FPS when the system had reached steady state and processed using custom MATLAB code derived from Refs.~\cite{wildeman2018real,moisy2009synthetic}. We observed the wave train to follow $A \propto r^{-1/2}$ in accord with established surface wave theory and follow the known dispersion relation for gravity-capillary waves:
\begin{equation}
    \label{GCDispersionRelation}
    \omega^2(k) = \left(gk+\frac{\gamma k^3}{\rho}\right)\tanh{\left(h_\text{rest}k\right)},
\end{equation}
where $\gamma$ is the fluid's surface tension, $k$ is the wavenumber, and $g$ is the standard gravity (see SI)~\cite{lamb1924hydrodynamics}.

Fast Checkerboard Demodulation analysis of steady-state waves between the boat and wall reveals a net field propagating outward from the boat (Fig.~\ref{fig:Waves&Imaging}(b)). These waves share $\omega$ with those emitted on the boat's far side and far from boundaries, but possess reduced $A$ regardless of $\omega$ (Fig.~\ref{fig:Waves&Imaging}(c)). We surmise that when the boat is sufficiently close to the wall, reflected waves return with non-negligible energy and modulate the free surface height at the hull. This modulation impedes concurrent wave generation on the side nearest the wall while minimally affecting the opposite side. Consequently, the steady-state amplitude between the hull and wall is reduced. We liken these dynamics to the reductions in height when jumping off a deformable medium~\cite{aguilar2016robophysical} or pumping a swing with poor timing~\cite{bae2005optimal}.

\begin{figure}[t!]
    \includegraphics[width=0.5\textwidth]{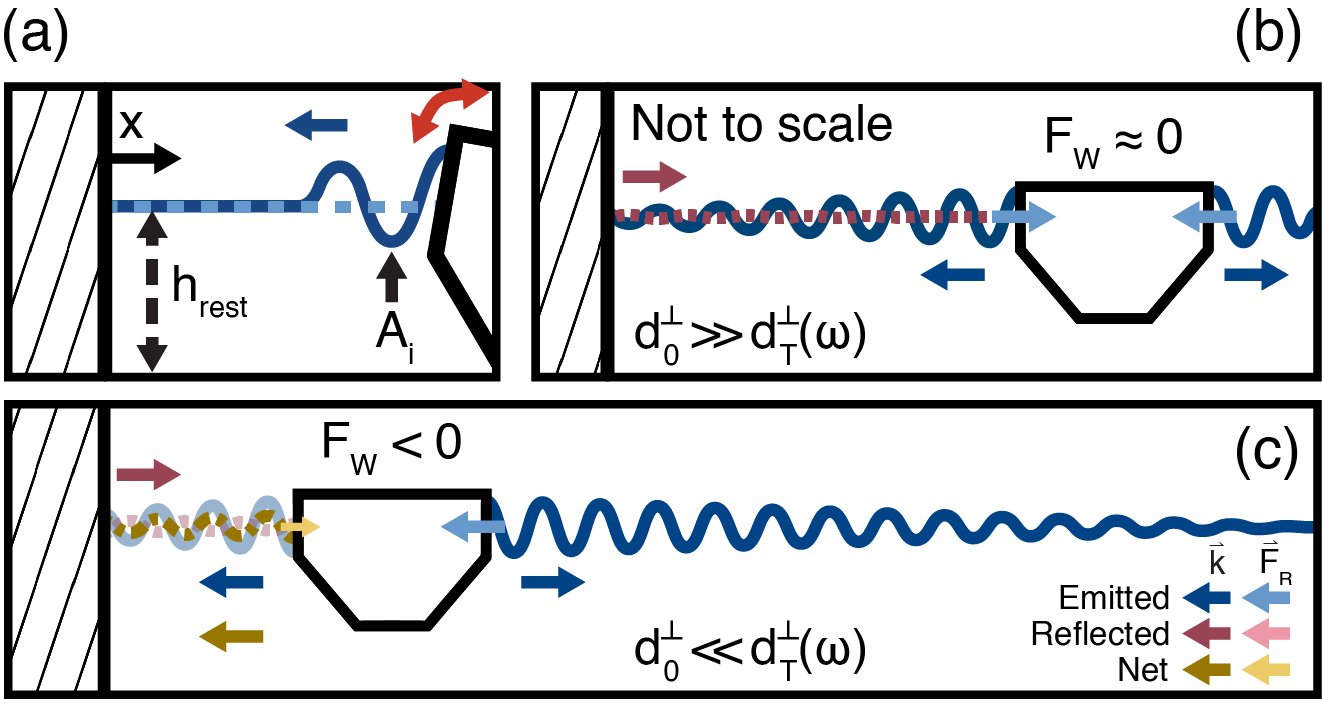}
    \centering
    \caption{\textbf{Hypothesis for attractive self-propulsion via Hocking field generation.} (a) Regardless of $d^\perp_0$, the boat initially generates a symmetric wave field. (b) When $d^\perp_0 \gg d^\perp_T$, the reflected waves have insufficient energy to affect the boat. (c) When $d^\perp_0 \ll d^\perp_T$, the reflected waves perturb the free-surface height at the boat, yielding a reduced-amplitude field. This amplitude asymmetry produces a net radiation force toward the boundary measured and predicted in Fig.~\ref{fig:PendulumExpts}.}
    \label{fig:CartoonHypotheses}
\end{figure}

\textit{A hydrodynamic model}~--~Armed with an understanding of the wave fields both near and far from boundaries, we motivate the boat's locomotive behavior as it relates to $d^\perp$  (Fig.~\ref{fig:CartoonHypotheses}). The existence of a radiative force incident on a wave emitter and proportional to square amplitude is a classical result~\cite{lamb1924hydrodynamics,longuet1964radiation} observed in many systems with asymmetric wave generation~\cite{denardo2009water,lee2017fluctuation,rhee2022surferbot}. When the boat is far from boundaries, the generated waves are spatially symmetric, leading to a net-zero $F_W$. Near boundaries, reflected waves induce an amplitude asymmetry, resulting in a finite $F_W$. We postulate that at a certain $d^\perp_T$,  the reflected wave will have insufficient energy to generate the asymmetric Hocking field. However, the reflected wave will not have dissipated enough for the boat to be considered far from boundaries; instead, the impingement of the reflected wave on the boat will lightly force it away from the wall.

Though the amplitude dynamics successfully describe the boat's attraction and motionlessness for small and large $d^\perp$ respectively, they provide insufficient reasoning for $F_W$'s observed frequency dependence. As indicated in Refs.~\cite{li1991,kardar1999,aminov2015}, such nonequilibrium amplitude dynamics will necessarily depend on specific details of the system. Therefore, we hypothesize that the unique properties of gravity-capillary waves are relevant to these complex hydrodynamics. Work by Hocking on the interactions of gravity-capillary waves with hard surfaces emphasizes the importance of wavenumber (alternatively frequency) to radiation and reflection~\cite{hocking1987reflection,hocking1987damping,hocking1988capillary}. Upon reflecting off a rigid boundary, gravity-capillary waves dissipate substantial energy through complex contact-line dynamics~\cite{michel2016acoustic}. Within the accessible wavenumber range for our boat, the reflection coefficient $R<0.22$ with $R \propto k^3$ and $k^{0.85}$ for $k\lesssim 7~\text{m}^{-1}$ and $k\gtrsim 20~\text{m}^{-1}$ respectively~\cite{hocking1987reflection}. Coupled with the aforementioned amplitude modulation, this wavenumber dependence suggests that higher frequency waves will have sufficient energy to induce attraction at farther hull-wall distances.

Further, gravity-capillary waves radiated by a vertically oscillating body have energy given by the following~\cite{hocking1988capillary}:
\begin{equation}
    E_R = \frac{\pi}{2}\left(1+\frac{3\gamma k^2}{\rho g}\right)A^2.
    \label{GCRadiationEnergy}
\end{equation}
Considering the boat as two back-to-back, semicircular wave emitters, this expression implies the following radiation force incident upon one side:
\begin{equation}
    |F_R(k)| = \frac{E_R k}{4\pi} = \left(\frac{k}{8}+\frac{3\gamma k^3}{8\rho g}\right)A^2\big(\omega(k)\big).
    \label{GCRadiationForce}
\end{equation}
The factor of $4\pi$ accounts for projecting the wave momentum normal to the semicircular boundary (see, e.g., Ref.~\cite{lee2017fluctuation} for a more detailed derivation). For our boat which has nontrivial $A(\omega)$ (Fig.~\ref{fig:Waves&Imaging}(c)), the predicted $F_R(\omega)$ follows a power law with exponent between $3$ and $4$ (Fig.~\ref{fig:Waves&Imaging}(d)). We reiterate that the amplitude measurements were taken within the attractive regime, and so we shift the origin of our power law to the observed threshold frequency for attraction. The difference between $F_R$ on either side of the boat yields a predicted $F_W$; this prediction matches well with experimental results without any fitting parameters (Fig.~\ref{fig:PendulumExpts}(d)).

We summarize our postulated model of the boat's boundary-driven locomotion in four regimes. In all cases, when the boat first emits waves, the field is symmetric, leading to a net-zero radiation force on the boat (Fig.~\ref{fig:CartoonHypotheses}(a)). When $d^\perp_0 \gg d^\perp_T$, the waves reflected off the boundary return to the boat with negligible energy compared to emission. Consequently, the boat experiences a force negligibly close to zero (Fig.~\ref{fig:CartoonHypotheses}(b)). When $d^\perp_0 \ll d^\perp_T$, the reflected waves hinder wave generation between the boat and wall, leading to an observed amplitude reduction. Meanwhile, waves on the far side remain unchanged; this broken symmetry yields a net force appearing as a boat-wall attraction (Fig.~\ref{fig:CartoonHypotheses}(c)). When approaching $d^\perp_T$ from $d^\perp_0 >d^\perp_T$, reflected waves have insufficient amplitude to affect wave generation but still carry non-negligible momentum. Symmetry is again broken and the boat experiences a slight repulsive force. Since the energy of a reflected gravity-capillary wave increases with $k$, $d^\perp_T$ also increases with $k$ (and, consequently, $\omega$). Should the original choice of $d^\perp_0$ be retained while increasing $\omega$, the reflected waves will then have sufficient energy to affect wave generation, causing the same result as when $d^\perp_0 \ll d^\perp_T$.

We note that our model can only explain the boat's steady-state position using wave amplitudes measured in that state. In 1D simulations of a free-floating boat that responds to the measured steady-state wave force and drag, the simulated boat always reaches the boundary faster than in experiment (see SI). An interesting notion is that the moving boat may experience a weaker wave force than in steady state due to the wave field's finite propagation time. In a dynamical system, the ever-changing boundary conditions may limit the extent to which the wave field can respond and evolve, leading to weaker transient fluctuation-induced forces. Additionally, Hocking's theories on gravity-capillary waves require both the emitter and reflecting boundary to be stationary on average. A much harder problem then is computing the system dynamics as the wave field updates; how would one compute the position versus time of the attracting boat in a dynamic environment? Indeed, we find the boat exhibits complex attractive modes like ``towing'' in response to a moving boundary (see SI, Movie S5). These dynamical experiments will help characterize transient locomotive states owed to Hocking fields in stationary and active environments.

\textit{Conclusion}~--~In this Letter, we revealed how a symmetrically oscillating robot can use confined hydrodynamic surface wave fields -- which we refer to as ``Hocking fields'' -- to locomote without the need for a traditional propulsion mechanism and made the first direct measurement of the corresponding force. In doing so, we add to the growing list of fluctuation-induced forces that employ surface wave fields both for propulsion and nonlocal interaction with fellow substrate occupants~\cite{wilcox1979sex,boersma1996maritime,buenzli2008,denardo2009water,hanke2013,parra2014casimir,ray2014casimir,ni2015tunable,aminov2015,harris2013wavelike,lee2017fluctuation,pucci2016nonspecular,roh2019honeybees,lee2019milli,saenz2021emergent,ho2021capillary,yuan2021wave,rhee2022surferbot,ko2022small,benham2022gunwale}. By symmetrically generating waves near a boundary, our boat takes advantage of the reflection dynamics unique to gravity-capillary waves to self-propel exclusively via wave drag with frequency- and distance-dependent locomotive modes. Our robophysical approach enables a convenient method to discover features of nonequilibrium, self-induced fluctuation-induced forces. The flexibility of this approach encourages future experiments that are not strictly limited to the fluid surface. Practically, manipulation of oscillation spectrum in response to transient conditions may prove valuable in expanding the range and strength of such interactions.

\begin{acknowledgments}
\textit{Acknowledgments}~--~We thank Enes Aydin for designing and constructing the tank apparatus and for helping during the boat design phase. We thank Ryan Hirsh for assisting with pendulum experiments. We thank Paul Umbanhowar for helpful comments and discussion. This work was funded by the Army Research Office Grants No. GR00008673, No. W911NF2110033 and a Dunn Family Professorship (D.I.G.).
\end{acknowledgments}

\section{Supplementary Information}
\renewcommand\thefigure{S\arabic{figure}}
\setcounter{figure}{0}

\begin{figure}[b!]
    \includegraphics[width=0.5\textwidth]{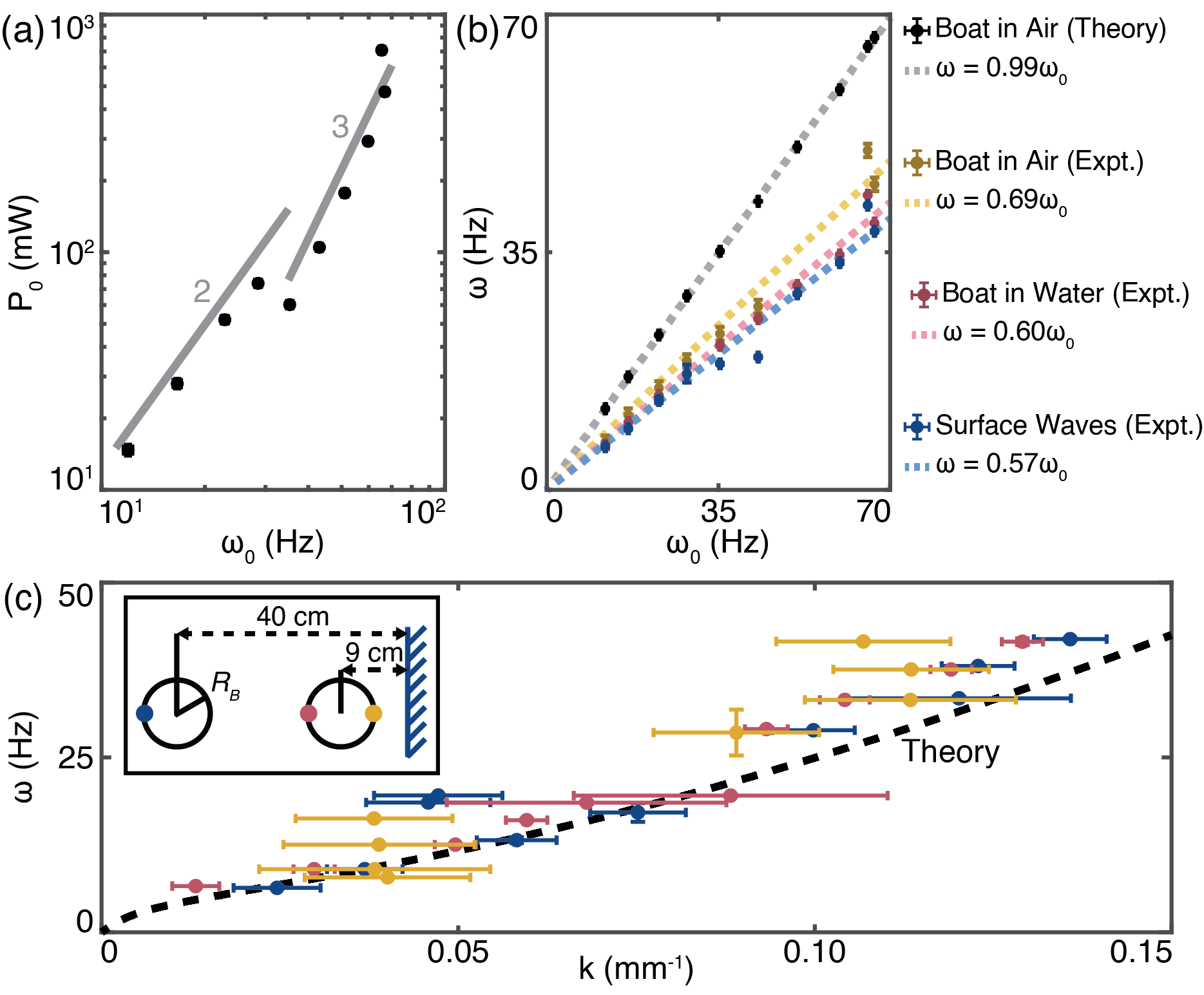}
    \centering
    \caption{\textbf{Eccentric motor oscillation drives boat vibration and consequent wave generation.} (a) Motor frequency response falls into two distinct modes with a resonance emerging at the crossover. (b) The boat's vibration is damped significantly due to coupling effects between the motor and boat and the fluid surface and hull. (c) Wave amplitude measurements yielded a dispersion relation comparable to established theory on gravity-capillary waves~\cite{lamb1924hydrodynamics}.}
    \label{fig:Motor}
\end{figure}

\textit{Details of boat vibration}~--~The robot boat vibrates in response to the oscillation of an internally mounted eccentric motor (Vybronics Inc. Cylindrical Vibration Motor VJQ24-35K270B). High-speed measurements of the motor in motion reveal a temporally consistent frequency $\omega_0$ in response to fixed power input $P_0$. We observe two response modes: $P_0 \propto \omega_0^2$ and $\omega_0^3$ for $\omega_0 \lesssim$ and $\gtrsim 30$ Hz respectively (Fig.~\ref{fig:Motor}(a)). The boat resonates at the crossover between modes, which appears experimentally as both the maximum in $A$-$\omega$ space (see main text) and the only significant deviation from linearity in $\omega$-$\omega_0$ space (Fig.~\ref{fig:Motor}(b)).

We cast the boat's vibratory response in 1D using the rotational analog of Newton's Second Law. By suspending the boat in midair on a string, we eliminate the need to model the complex feedback mechanisms owed to surface wave generation. Consequently, the relevant torques on the boat hull are produced by gravity, air drag, and the eccentric motor, which we model as a rotating unbalance~\cite{palm2014system}:
\begin{equation}
    \newcommand{\sgn}{\operatorname{sgn}}
    \label{BoatVibrationEoM}
    \begin{split}
        I\Ddot{\theta} = m_0\epsilon_0\omega_0^2R_0\sin{\left(\omega_0 t-\theta\right)}-mgR_B\sin{\theta} \\
        -\frac{\pi}{10}\rho c_D R_B^5 \sgn\big(\Dot{\theta}\big)\Dot{\theta}^2,
    \end{split}
\end{equation}
where $I$, $m$, $R_B$ and $c_D$ are the boat's moment of inertia, mass, radius, and drag coefficient respectively; $m_0$ and $\epsilon_0$ are the rotating unbalance's mass and radius respectively; $R_0$ is the distance between the motor shaft and boat hull; and $\rho$ is the density of air. We simulate Eq.~\eqref{BoatVibrationEoM} with an ordinary differential equation solver in MATLAB and find an expected response frequency $\omega = 0.99\omega_0$ (Fig.~\ref{fig:Motor}(b)). However, physical measurement of the boat's in-air vibration reveals a reduced response driven at 69\% the motor's frequency. Placed in water, the boat's vibration drops further to 60\%, with the surface wave frequency nearby at 57\%. We attribute these discrepancies to two sources of damping, namely the motor's non-idealized mounting to the boat and the coupling between the fluid surface and the hull.

\begin{figure}[b!]
    \includegraphics[width=0.5\textwidth]{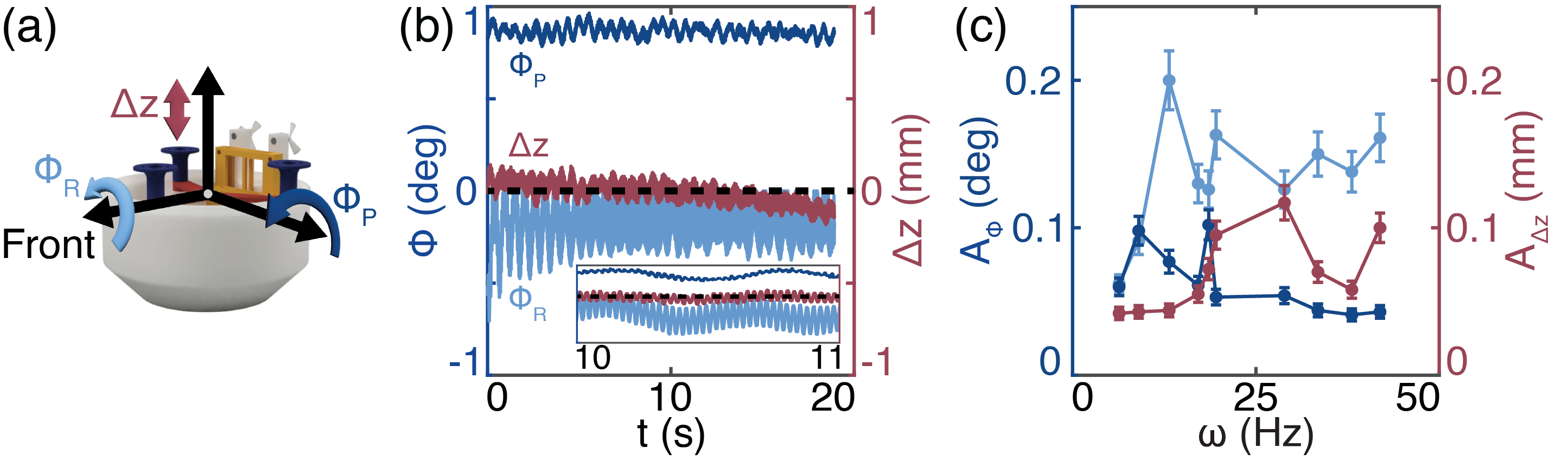}
    \centering
    \caption{\textbf{Boat vibrates primarily along the roll axis with small amplitude.} (a) Diagram of roll, pitch, and vertical displacement axes. (b) Archetypal boat vibration at 38.1 Hz. (c) Amplitudes of oscillation along roll, pitch, and vertical displacement axes. For nearly all accessible frequencies, the primary boat oscillation occurs along the roll axis.}
    \label{fig:Optitrack}
\end{figure}

To better understand the boat's vibratory response in 3D, we tracked the oscillation with multiple high-speed cameras (OptiTrack) at 360 FPS (Fig.~\ref{fig:Optitrack}). Unlike many established systems that employ periodic heaving (vertical) motions to generate surface waves~\cite{tatsuno1969transfiguration,hulme1982wave,hocking1988capillary,taneda1991visual,benham2022gunwale}, our boat undergoes minimal vertical displacement. Instead, the eccentric motor induces oscillations primarily along the fore-aft (roll) axis. Still, the boat's overall vibrational motion is miniscule, with a maximum roll amplitude $\phi_R = 0.20\degree \pm 0.02\degree$ corresponding to a vertical amplitude of $0.21 \pm 0.02$ mm.

\textit{Surface currents}~--~As a further check on our hypothesis that the boat's attraction toward and repulsion from boundaries is the result of surface waves, we investigated the possible existence of surface currents induced by the boat's wave generation. After fixing the boat's lateral position such that vibration was unimpeded, we seeded the fluid surface with lycopodium powder (CAS number: 8023-70-9) for use with open-source Particle Image Velocimetry (PIV) software in MATLAB~\cite{thielicke2021particle}. Results are shown in Movie S4.

\begin{figure}[t!]
    \includegraphics[width=0.5\textwidth]{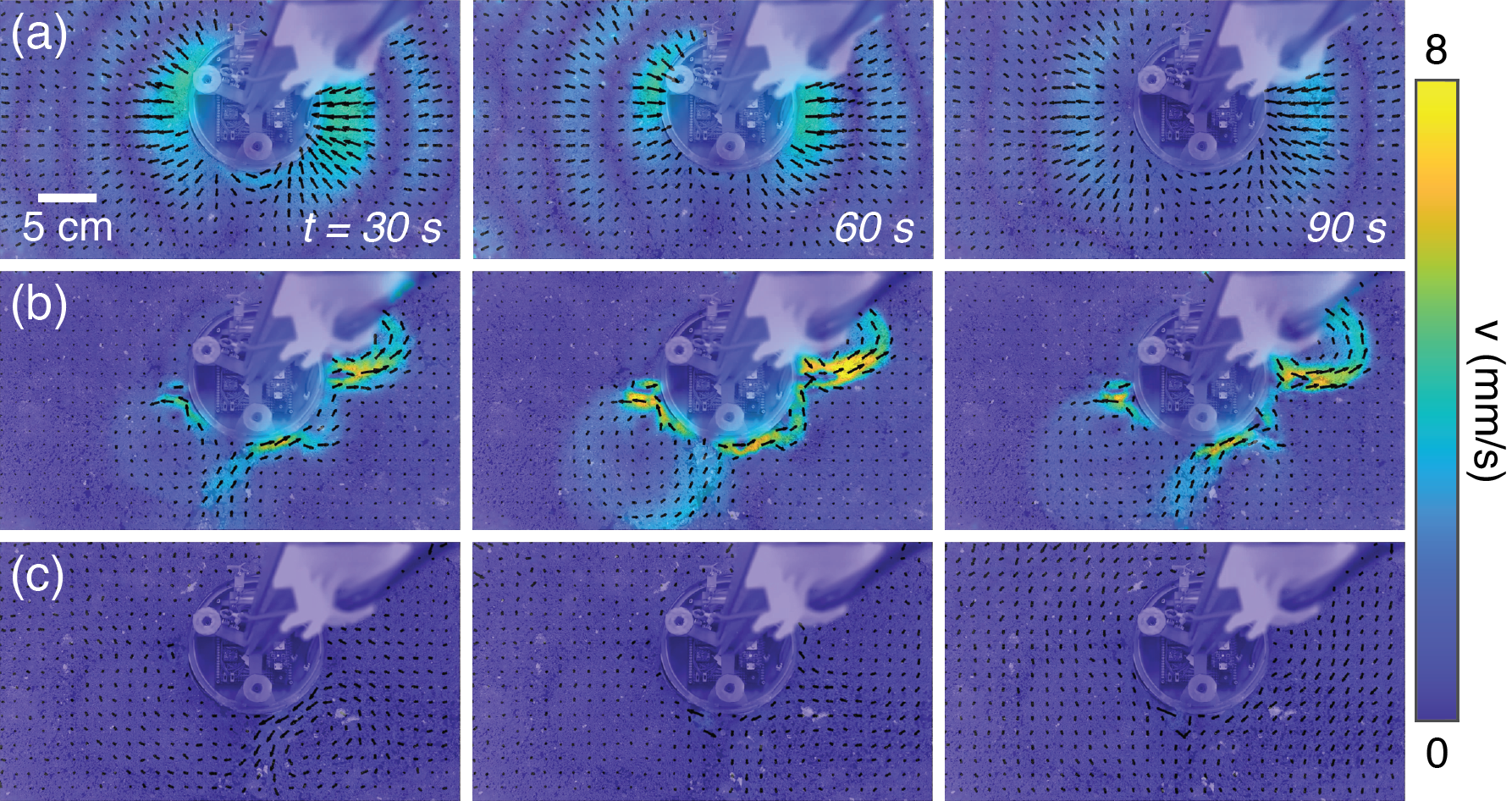}
    \centering
    \caption{\textbf{Boat vibration generates minimal surface currents.} (a-c) Surface currents induced by wave generation at 6.3, 19.6, and 33.5 Hz respectively. Currents shown are representative of behaviors within the three distinct regimes: $\omega \in (0,19), [19,29],$ and $(29,42]$ Hz.}
    \label{fig:PIV}
\end{figure}

For $\omega < 19$ Hz, no surface currents are produced, and seed particles trace circular paths in the vertical plane as they bob over the waves (Fig.~\ref{fig:PIV}(a)). For $19$ Hz $\leq \omega \leq 29$ Hz, vortices emerge as seed particles are drawn in at the fore and aft, circulate along the boat perimeter, and eject in jets at the left and right sides with maximum velocity $v \approx 8$ mm/s (Fig.~\ref{fig:PIV}(b)). We rationalize the ingress and egress positions as the locations with the weakest and strongest vibratory motion respectively, a result of the orientation of the eccentric motor driving the oscillation. For $\omega > 29$ Hz, a few smaller vortices with maximum velocity $v \approx 1$ mm/s emerge inconsistently around the boat perimeter (Fig.~\ref{fig:PIV}(c)). All three regimes persist when the boat is brought near a boundary.

The frequency dependence of these distinct surface current modes does not correlate with that of Hocking fields as described in the main text. Furthermore, when orienting the boat such that the primary surface jets expel toward a nearby boundary, the boat still experiences an attractive force where mechanical intuition suggests a repulsion should emerge. For these reasons, we rule out surface currents as a probable cause for Hocking fields and reaffirm our surface wave hypothesis.

\textit{Synthetic Schlieren imaging}~\cite{SchlierenInCentipedesNote}~--~Before starting the experiments, we captured a reference image of the background pattern (checkerboard) as seen through a still free surface with the high speed camera. During experiments, surface waves appeared as a distortion field $\overrightharp{u}$ applied to the checkerboard. We compared the spatial Fourier transform of the distorted checkerboard to that of the reference image to find how the carrier peaks were modulated. When the free surface curvature had focal length greater than the distance to the background pattern (i.e., the invertibility condition is met~\cite{moisy2009synthetic}) we filtered the modulated signal to extract $\overrightharp{u}(t,\overrightharp{r})$, which is proportional to the gradient of the free surface height. Moisy and colleagues quantify this invertibility condition as follows:
\begin{equation}
    \label{InvertibilityCondition}
    h_p < h_{p,c} = \frac{\lambda^2}{4\pi^2\alpha\eta_0},
\end{equation}
where $h_p$ is the effective surface-pattern distance, $h_{p,c}$ is the free surface focal length, $\lambda$ is the wavelength, $\alpha$ is the ratio of indices of refraction given by $1-n_\text{air}/n_\text{fluid}$, and $\eta_0$ is the wave amplitude~\cite{moisy2009synthetic}. Further, we adapted the open-source code in Ref.~\cite{wildeman2018real} for use with our apparatus, incorporating a scale factor to account for additional interfaces between the background pattern and the fluid free surface~\cite{moisy2009synthetic}.

\begin{figure}[t!]
    \includegraphics[width=0.5\textwidth]{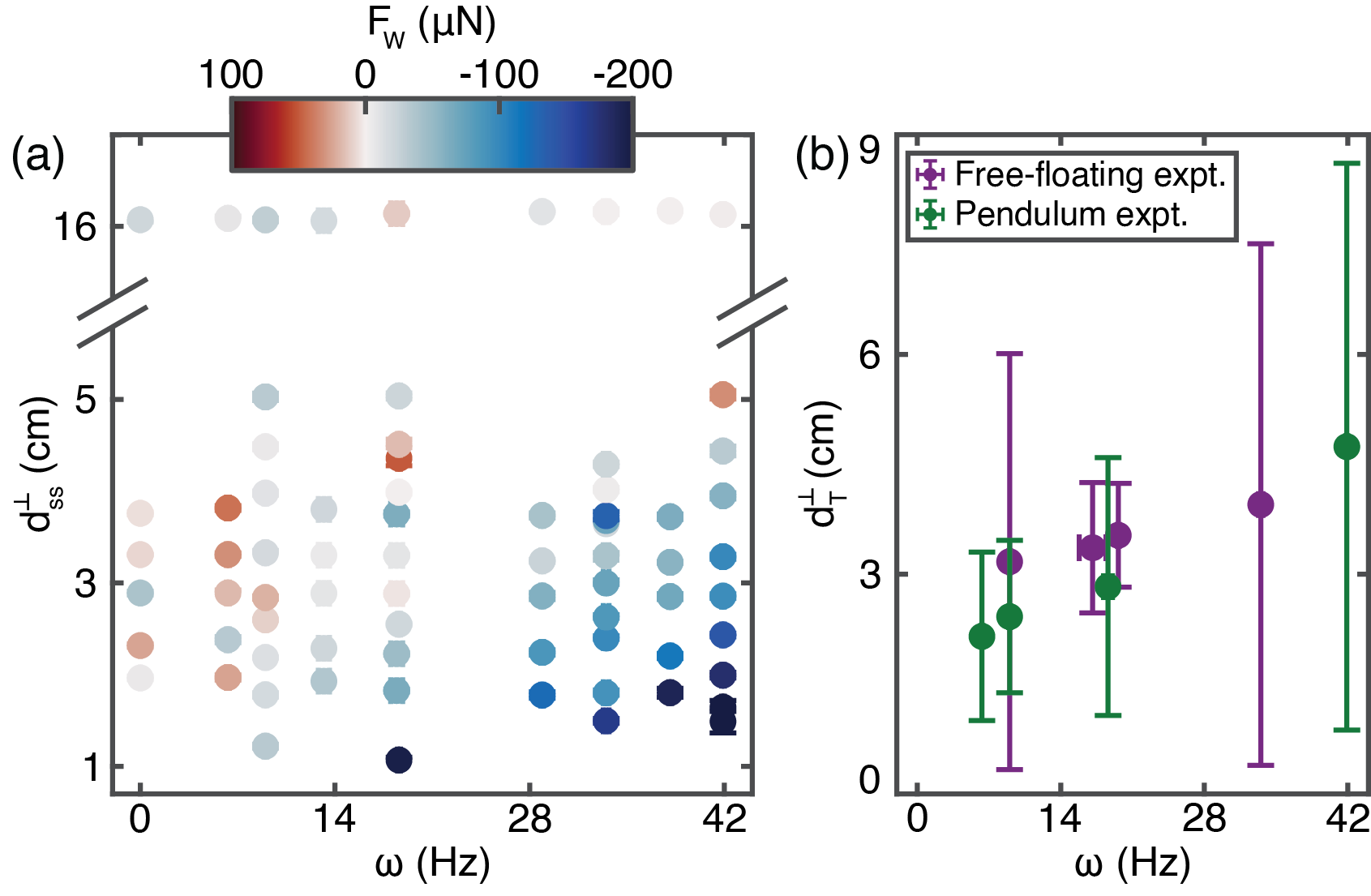}
   
    \centering
    \caption{\textbf{Heatmap of near-boundary propulsive force measurements and corresponding estimate of attraction-repulsion threshold}. (a) Simultaneous dependence of $F_W$ on both $d^\perp_\text{ss}$ and $\omega$ where each point corresponds to the average of 3 trials. As expected, the $F_W$ heatmap is qualitatively similar to the $\ddot{d}^\perp$ heatmap in the main text. (b) We estimate $d_T^{\perp}(\omega)$ using heatmap data from both free-floating and pendulum experiments. Parameter choices near boundaries that lie below and above the $d_T^{\perp}(\omega)$ curve should result in attraction and repulsion respectively.}
    \label{fig:ForceHeatmap}
\end{figure}

Quantitatively identifying where the invertibility condition fails requires knowledge of wave properties that are not known a priori and cannot be reliably obtained from the reconstruction itself. However, we note that failed reconstruction surface height data typically is highly discontinuous, both with itself and with successfully reconstructed surface heights. We used this characteristic to estimate regions where the reconstruction failed per video frame with an autocorrelation method described by the following steps:
\begin{enumerate}
    \itemsep0em 
    \item Perform a 2D spatially-moving variance with square kernel given by the 8-way nearest pixel neighbors.
    \item Compare the moving variance to a threshold value. We obtained our threshold through trial-and-error but postulate that it is related to the effective distance between the free surface and background pattern.
    \item Convert any pixels for which the variance exceeds the threshold to a mask.
    \item Perform minor cleanup on the mask using morphological operations. The result is an estimate of all failed surface reconstructions in the frame.
\end{enumerate}

\textit{Probing a simplified force model in simulation}~--~We probe a simplified force model to describe the boat's position in 1D in response to attraction. The two relevant forces are the wave force $F_W$ and drag $F_D$. We observed in our pendulum experiments that for a fixed wave frequency $\omega$, $F_W$ is linear with $d^{\perp}$. We select an arbitrary $\omega$ and use a linear fit to model the wave force $F_W = \alpha d^{\perp}+\beta$, where $\alpha$ and $\beta$ are fitting parameters.

During free-floating experiments near a boundary, we observed the boat's mean speed per trial to have a minimum of 0.3 mm/s. We also observed the boat to have a top speed of 30 mm/s. Taking the boat's characteristic length $L$ to be its diameter, we estimate the fluid's Reynolds number $Re = \rho v L/\mu$ to range from 40 to 4000 during free-floating experiments, where $\rho$ and $\mu$ are the fluid's density and dynamic viscosity respectively. This range indicates that the boat's translational motion is dominated by inertia; we consequently model drag using the equation $F_D = \rho A c_D (\dot{d}^{\perp})^2/2$, where $A$ is the boat's cross-sectional area submerged in the fluid and $c_D$ is the coefficient of drag. We observed a good fit between position data from 1D coasting experiments and inertial drag predictions. Further, we measured $c_D = 1.12\pm0.20$ for our boat, which has a cylindrical hull with a frustum underside. We expect the frustum to reduce drag slightly compared to a perfect cylinder. Indeed, our measured drag coefficient is slightly lower than the drag coefficient for a cylinder $c_D = 1.17$~\cite{Hoerner1965}. Taken together, we use the following as our simplified force model for attraction: $m\Ddot{d}^{\perp}=F_W(d^{\perp})-F_D\big((\dot{d}^{\perp})^2\big)$.
        
\begin{figure}[t!]
        \includegraphics[width=0.5\textwidth]{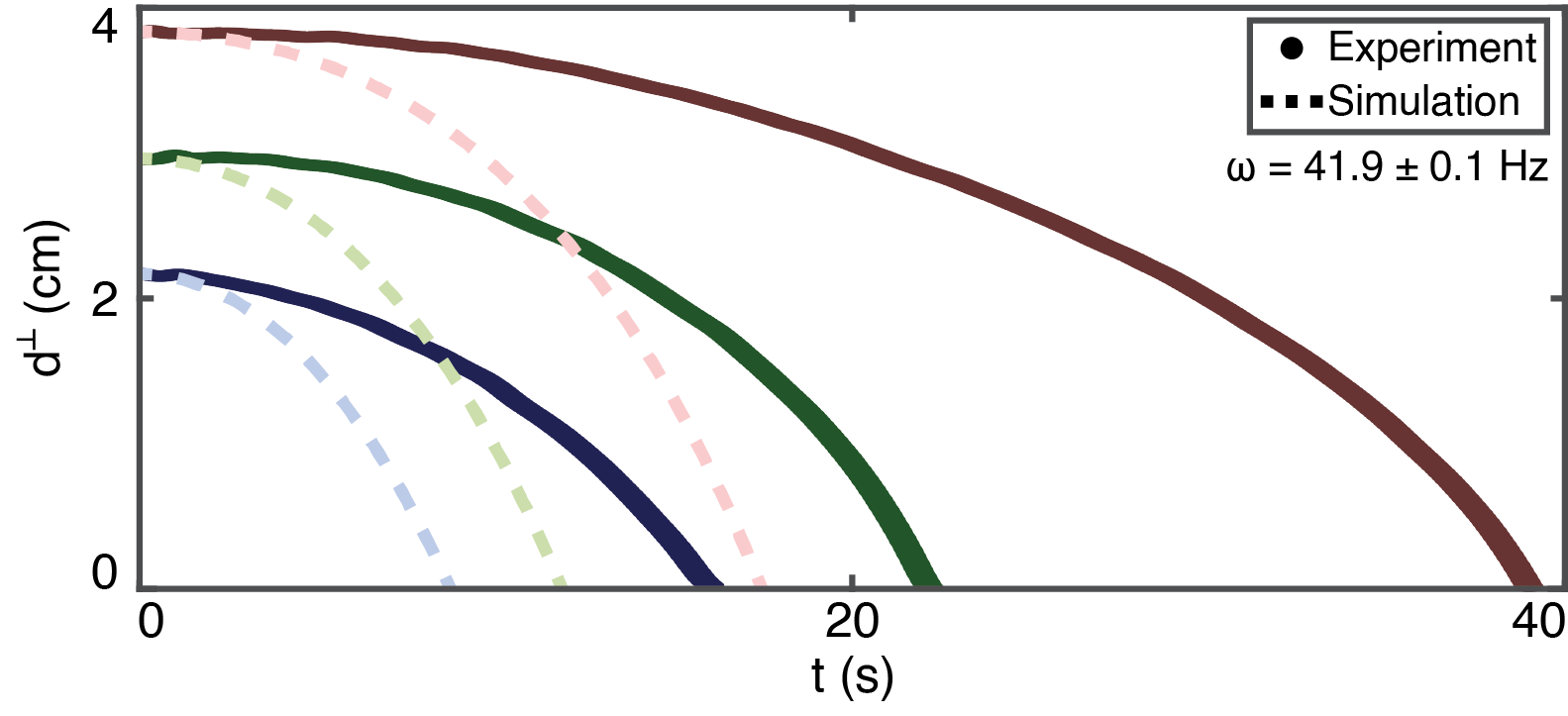}
        \centering
        \caption{\textbf{Simplified force model predicts collision with boundary twice as fast as in experiment.} Archetypal comparisons between simulated force model and experiment.}
        \label{fig:Sim}
    \end{figure}

In simulating this model with an ordinary differential equation solver in MATLAB, we observe qualitative similarity between experimental and theoretical trajectories. However, across a variety of wave frequencies and initial positions within the attractive regime, the simulated boat always reaches the boundary faster than in experiment by a factor of $1.9\pm0.4$ (Fig.~\ref{fig:Sim}). This consistent discrepancy suggests two possibilities: (1) our model neglects some relevant drag term, and (2) the steady-state wave force measured with pendulum experiments is stronger than the transient wave force experienced by a moving boat. Given how slow the boat's translational speed is, we doubt there is a relevant drag term that would fully account for this large discrepancy. More interesting is the notion that the moving boat experiences a weaker wave force. There may be some finite propagation time for the wave field to fully establish itself and induce the force we observed in our steady-state pendulum experiments. In a dynamical system, the ever-changing boundary conditions may limit the extent to which the wave field can respond and evolve, leading to weaker transient fluctuation-induced forces. As noted in the main text, more work is needed to fully understand the implications of these complex dynamical states.

\begin{figure}[t!]
    \includegraphics[width=0.5\textwidth]{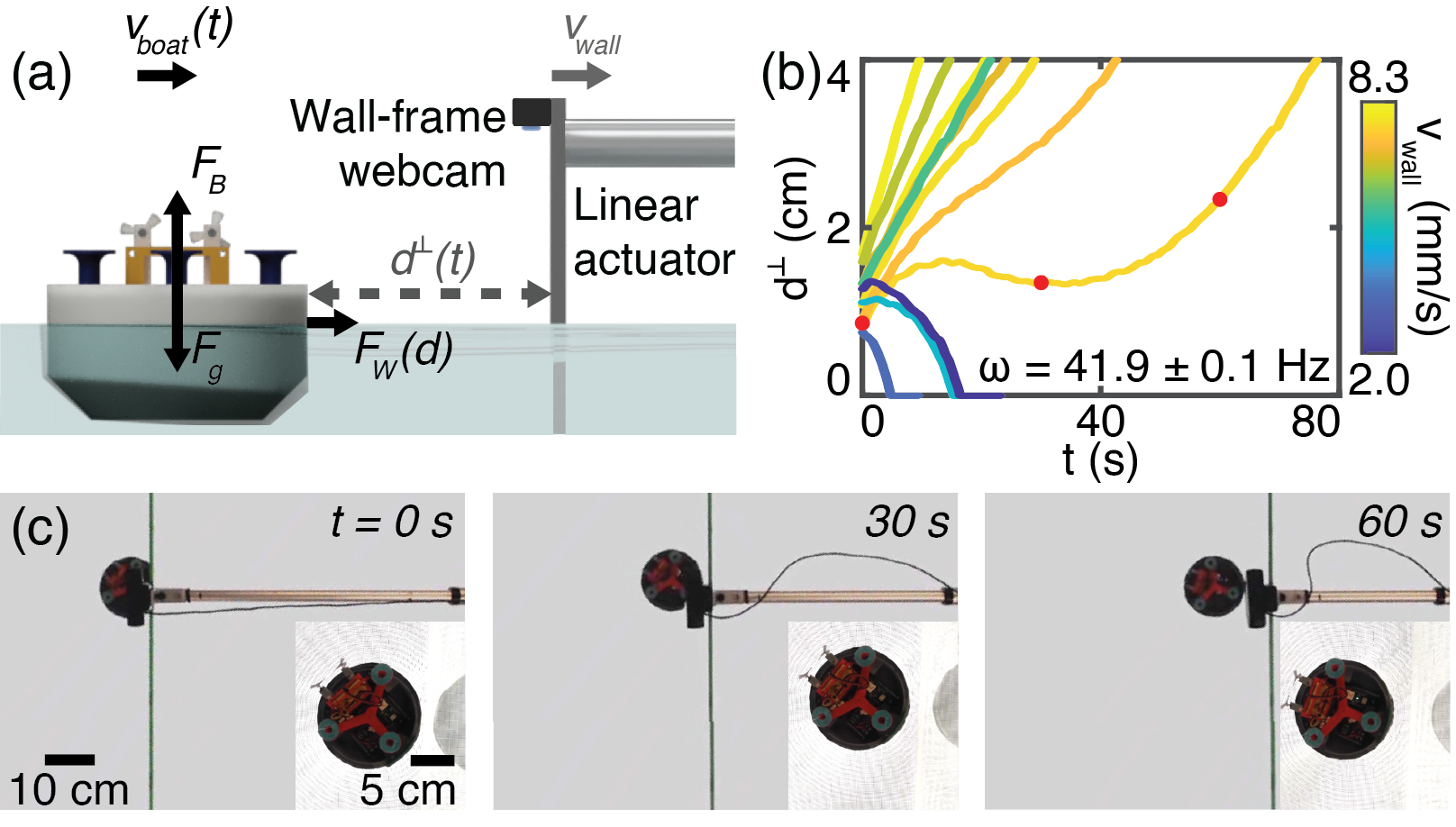}
    \centering
    \caption{\textbf{Attractive Hocking field enables towing by moving boundary.} (a) Force diagram for towing experiments. (b) $d^\perp$ versus $t$ with $t_0$ corresponding to the wall's initialization. For low $v_\text{wall}$, the boat catches the wall within 20 s. For high $v_\text{wall}$, the wall rapidly outpaces the boat. Slight variance of $d^\perp_0$ caused a non-monotonic trend with $v_\text{wall}$. (c) Lab- and (\textit{Inset}) wall-frame time series of towing experiment closest to $v_\text{wall}$ bifurcation. Snapshots correspond to red points in (b).}
    \label{fig:TowingExpts}
\end{figure}

\textit{Probing response to moving boundaries}~--~Though the constrained pendulum system enabled measurement of the Hocking field's corresponding radiation force, it restricted phenomenological exploration to a firmly asymmetric-field regime. To probe the existence of transition dynamics between the asymmetric and symmetric regimes of the Hocking field phenomenon, we measured the response of a free-floating boat with constant $\omega$ to a wall retreating with constant speed $v_\text{wall}$ in 1D (Fig.~\ref{fig:TowingExpts}, Movie S5). When initiated with a boat-wall attraction, we posit the existence of a bifurcation in $v_\text{wall}$ at which the wall would ``tow" the boat with constant $d^\perp$.

We mounted the wall from previous experiments on a linear actuator (Firgelli\textsuperscript{\textregistered} FA-240-S-12-18) powered with constant current and pulse-width modulated voltage. Velocimetry measurements~\cite{thielicke2021particle} revealed minimal surface currents ($v< 5$ mm/s) near the wall's center; consequently, we performed experiments in this central region. We chose $d^\perp_0 = 1.18 \pm 0.28$ cm and $\omega = 41.9$ Hz such that the boat started firmly within the attractive regime. Once $v_\text{boat} \approx v^\text{min}_\text{wall} = 2$ mm/s, we initialized the wall and recorded $d^\perp(t)$ using an onboard Logitech C920 webcam (Fig.~\ref{fig:TowingExpts}(b)).

When $v_\text{wall} \leq 4.9$ mm/s, acceleration induced by the Hocking field was sufficient for the boat to catch the retreating boundary. For large $v_\text{wall}$, the wall swiftly outperformed the boat's locomotion. We most closely approached our expected bifurcation when $v_\text{wall} = 7.7$ mm/s. During this trial, the wall towed the boat with $d^\perp < 2$ cm for a total distance of $43$ cm. Despite the stringent boundary requirement for Hocking fields to appear, proper choice of $\omega$ and $v_\text{wall}$ enabled the boat to travel over 10x further than the corresponding $d^\perp_T(\omega)$.

\begin{figure}[t]
    \includegraphics[width=0.5\textwidth]{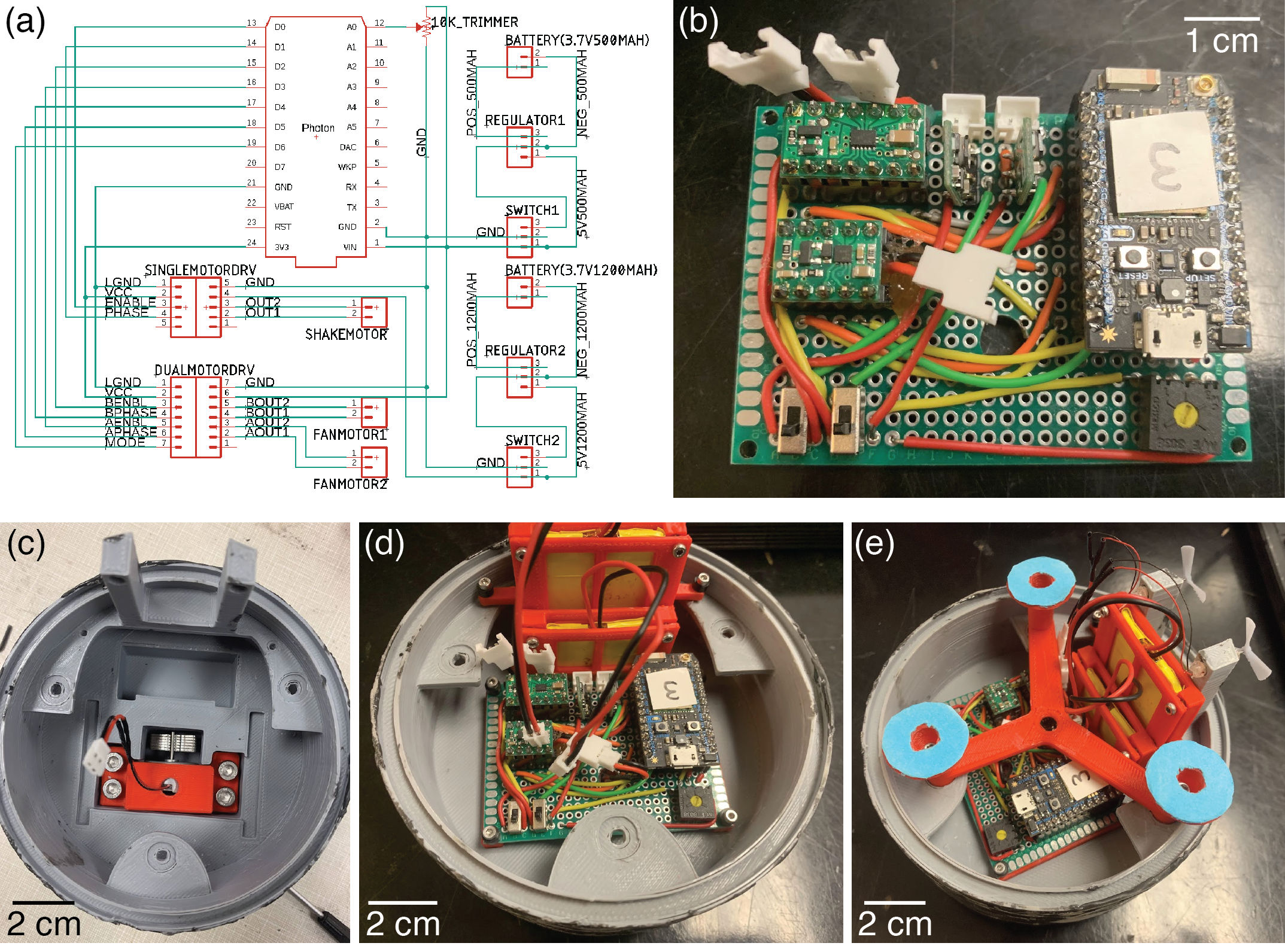}
    \centering
    \caption{\textbf{Robot boat assembly.} (a) Boat circuit board schematic. (b) Fully assembled circuit board. (c) Boat chassis with eccentric motor installed. (d) Boat chassis with circuit board and batteries installed. (e) Fully assembled boat.}
    \label{fig:BoatAssembly}
\end{figure}

\textit{Robot boat construction}~--~Parts for the robot boat are included in Tables~\ref{PurchasedPartsList} and~\ref{3DPPartsList}. Start by assembling the circuit board (Fig.~\ref{fig:BoatAssembly}(a-b)). We encourage hand-soldering components onto a perfboard because the boat's intense vibrations may rip small traces off a printed circuit board. Drill a small hole through the board for the shaker motor wires. Mount the circuit board to the shims through the guide holes using four M2 x 8mm screws with one spacer each. Solder charging cables onto each of the three motors for later use.

Waterproof the boat chassis exterior with an even layer of marine epoxy such that no printed plastic is visible. Once the coat cures, smooth out any rough patches with fine-grit sandpaper without revealing the plastic underneath. Mount four M5 embedment nuts in the guide holes at the bottom of the boat interior. Mount the eccentric motor in the semi-cylindrical cutout between the nuts. Orient the motor such that the wires are facing up and the motor is flush with the back wall. While holding the motor in place, thread the wires through the central hole in the eccentric motor clamp and tighten the clamp with four M5-0.8 x 16 mm screws (Fig.~\ref{fig:BoatAssembly}(c)).

\begin{table}[t]
    \centering
    \caption{\textbf{List of purchased parts for robot boat.}}
    \begin{tabular}{| p{0.5\linewidth} | p{0.5\linewidth} |}
    \hline
        \textbf{Part name} & \textbf{Brand \& Part No.}\\\hline
        Marine epoxy (\textbf{x1}) & J-B Weld Part No. 8271\\\hline
        Machine screws & \#4-40 x 1" (\textbf{x3})\newline M2 x 8mm (\textbf{x6})\newline M2 x 16mm (\textbf{x6})\newline M5-0.8 x 16mm (\textbf{x4})\\\hline
        Machine screw hex nuts  & \#4-40 x 3/32" x 1/4" (\textbf{x3})\newline M2-0.4 x 4mm x 1.6mm (\textbf{x2})\\\hline
        Brass knurled insert embedment nuts & M2 x 4mm x 3.5mm (\textbf{x6})\newline M5 x 6mm x 7mm (\textbf{x4})\\\hline
        Eccentric motor (\textbf{x1}) & Vybronics Inc. Part No. VJQ24-35K270B\\\hline
        Fan motor (\textbf{x2}) & uxcell Part No. 412 4x12mm\\\hline
        Single motor driver (\textbf{x1}) & Pololu Part No. 2990\\\hline
        Dual motor driver (\textbf{x1}) & Pololu Part No. 2135\\\hline
        5V step-up voltage regulator (\textbf{x2}) & Pololu Part No. 2564\\\hline
        Mini slide switch (\textbf{x2}) & Pololu Part No. 1408\\\hline
        3.7V 500mAh LiPo battery (\textbf{x1}) & Adafruit Part No. 1578\\\hline
        3.7V 1200mAh LiPo battery (\textbf{x1}) & Adafruit Part No. 258\\\hline
        100mAh LiPo USB charger (\textbf{x2}) & Adafruit Part No. 1304\\\hline
        Female headers (\textbf{x6}) & Adafruit Part No. 2886\\\hline
        Wire-to-board pin header (\textbf{x2}) & Newark Part No. B2B-PH-K-S(LF)(SN)\\\hline
        10k$\Omega$ 0.5W trimmer (\textbf{x1}) & Bourns Inc. Part No. 3386P-1-103LF\\\hline
        3.7V male-female charging cables (\textbf{x3}) & NIDICI 1S\\\hline
        WiFi module (\textbf{x1}) & Particle Photon\\\hline
        Double-sided PCB board (\textbf{x1}) & 5cm x 7cm\\\hline
        Jumper wires (\textbf{x22}) & Assorted lengths\\\hline
    \end{tabular}
    \label{PurchasedPartsList}
\end{table}

\begin{table}[t]
    \centering
    \caption{\textbf{List of 3D-printed parts for robot boat.} All CAD files are available through Autodesk Fusion 360.}    
    \begin{tabular}[c]{| c | c |}
    \hline
        \multicolumn{2}{|c|}{\textbf{Printed in PLA at 0.20mm layer height}}\\\hline
        \textbf{Part name} & \textbf{Link to CAD file}\\\hline
        Boat chassis (\textbf{x1}) & \url{https://a360.co/3OYDJQJ}\\\hline
        Centering device (\textbf{x1}) & \url{https://a360.co/3OYDJQJ}\\\hline
        Marker platform (\textbf{x3}) & \url{https://a360.co/3OYDJQJ}\\\hline
        Battery compartments (\textbf{x1 each}) & \url{https://a360.co/3OYDJQJ}\\\hline
        Circuit board shims (\textbf{x1 each}) & \url{https://a360.co/3OYDJQJ}\\\hline
        Eccentric motor clamp (\textbf{x1}) & \url{https://a360.co/3FeCqZA}\\\hline
        \multicolumn{2}{|c|}{\textbf{Printed in resin at 0.050mm layer height}}\\\hline
        \textbf{Part name} & \textbf{Link to CAD file}\\\hline
        Fans (\textbf{x1 each}) & \url{https://a360.co/3yeETSg}\\\hline
        Spacer (\textbf{x6}) & \url{https://a360.co/37lFDu0}\\\hline
    \end{tabular}
    \label{3DPPartsList}
\end{table}

Mount six M2 embedment nuts in the guide holes on the largest battery compartment piece. Place the 1200mAh LiPo battery into that piece, then top it with the corresponding cover. Similarly, place the 500mAh LiPo battery into the small rear compartment piece and top with the small cover. Stack the two batteries and screw the entire assembly together with four M2 x 16 mm screws and two M2 x 8 mm screws. The covers will deform slightly to clamp the batteries in place as the screws are tightened.

Mount the battery compartments onto the boat (Fig.~\ref{fig:BoatAssembly}(d)). Place two M2 x 16 mm screws with one spacer each through the two holes in the battery compartment wings. Run these screws through the corresponding holes in the boat's back ledge. To rigidly mount the batteries, add an M2 nut to each screw and tighten until the nuts are snug with the underside of the ledge. Thread the eccentric motor wires through the circuit board’s central hole. Mount the circuit board to the boat by press-fitting the shims into their guide holes. Plug the eccentric motor into its corresponding connector on the circuit board. Do not plug in the batteries until use, but note that the 500mAh and 1200mAh batteries will eventually plug into the port near the WiFi module and the port near the motor drivers respectively.

Attach colored markers to the tops of the marker platforms as needed. Connect the centering device and marker platforms to the boat using three \#4-40 x 1” screws. To rigidly mount, add a \#4-40 nut to each screw and tighten until the nuts are snug with the undersides of the ledges and the marker platforms resist rotation.

Mount the fan motors snugly in their corresponding holes such that the wire side of the motor is flush with the interior side of the hole. Apply a small dot of hot glue to the wire side of the motors without blocking the battery compartments from being removed. Plug the fan motors into their corresponding connectors. Apply a small dot of superglue to the ends of both motor axles. Slide the fans onto the ends of the motor axles. The completed boat is shown in Fig.~\ref{fig:BoatAssembly}(e).

If the boat floats askew, weights can be added to the boat gunwale's inner lip. With the boat at rest in a water bath, place a small 2D bubble level on the centering device. Using the bubble level as a guide, attach weights to the inner lip with hot glue. Repeat as needed until the boat is level when floating.

\begin{figure}[t]
    \includegraphics[width=0.5\textwidth]{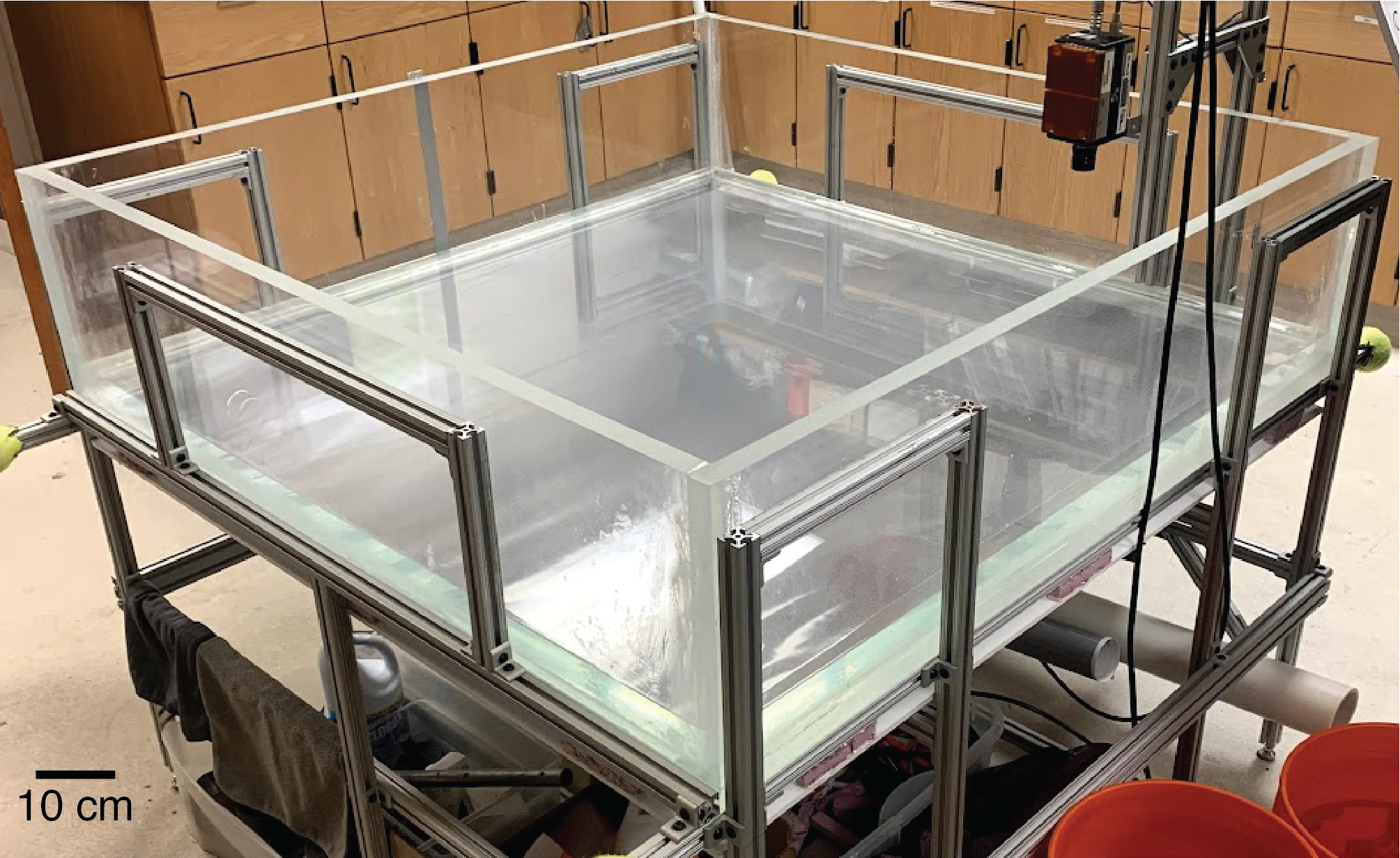}
    \centering
    \caption{\textbf{Fully assembled tank apparatus.}}
    \label{fig:TankAssembly}
\end{figure}

\begin{table}[t]
    \centering
    \caption{\textbf{List of purchased parts for tank apparatus.}}
    \begin{tabular}{| p{0.5\linewidth} | p{0.5\linewidth} |}
    \hline
        \textbf{Part name} & \textbf{Brand \& Part No.}\\\hline
        Aquarium-grade acrylic (\textbf{x5}) & Acrylite GP custom-cut sheets\\\hline
        Acrylic welding adhesive (\textbf{x3}) & SCIGRIP \#40\\\hline
        Clear silicone sealant (\textbf{x1}) & 3M Part No. 08661\\\hline
        Clear rubber sealant (\textbf{x1}) & Flex Seal aerosol spray\\\hline
        White LED panel (\textbf{x2}) & Metalux 2' x 4', 4700 lumens\\\hline
        T-slotted framing rails, structural brackets, \& fasteners & McMaster-Carr (\textbf{as needed})\\\hline
        Threaded leveling mounts & McMaster-Carr (\textbf{as needed})\\\hline
    \end{tabular}
    \label{TankPartsList}
\end{table}

\textit{Tank apparatus construction}~--~Parts for the tank apparatus are included in Table~\ref{TankPartsList}. Before building the tank, thoroughly deburr all acrylic sheet edges. Fuse the four acrylic wall pieces together with the welding adhesive. Once the walls are solidly together, fuse them as one piece with the acrylic base. Apply a layer of silicone sealant to all interior joints and a layer of rubber sealant to all exterior joints. Repeat as needed. After the sealants cure, flip the tank upside down and remove any dust or debris. Attach a checkerboard~\cite{wildeman2018real} to the bottom of the tank, evacuating air bubbles as needed. Right the tank and install LED panels underneath to backlight the checkerboard.

If needed, assemble an aluminum frame to house the tank and LED panels. The tank should be fully supported along its perimeter such that the underside lighting is unobscured. Level the frame using adjustable leveling mounts before starting any experiments. The completed tank apparatus is shown in Fig.~\ref{fig:TankAssembly}.


\begin{thebibliography}{10}
\bibitem{NegligibleVertDispNote}
The pendulum angle $\theta$ is related to the boat's horizontal displacement $\Delta x$ by the length of the fishing line $L$: $\sin{\theta}=\Delta x/L$. The boat's vertical displacement then must be $\Delta y = L(1-\cos{\theta}) = L(1-\sqrt{1-\Delta x^2/L^2})$. Typical values of $\Delta x$ were no larger than $3$ mm and $L=1.4$ m; therefore, the maximum vertical displacement is $3~\mu$m. Considering the order of the boat's radius $\orderof(R_B) = 10^{-2}$ m and emitted wave amplitude $\orderof(A) = 10^{-4}$ m, we assume a vertical displacement on the order $\orderof(\Delta y) \leq 10^{-6}$ m to be negligible.
\end{thebibliography}

\end{document}